\documentclass[12pt,epsf]{iopart}

\usepackage{graphicx}
\usepackage{amssymb,latexsym}

\newcommand{\beq}{\begin{equation}}
\newcommand{\beqa}{\begin{eqnarray}}
\newcommand{\eeq}{\end{equation}}
\newcommand{\eeqa}{\end{eqnarray}}
\newcommand{\lmk}{\left(}  \newcommand{\rmk}{\right)}
\newcommand{\lkk}{\left[}

\newcommand{\vp}{\varphi}
\newcommand{\simg}{\gtrsim}
\newcommand{\siml}{\lesssim}

\begin{document}

\title{Initial Conditions for Vector Inflation}

\author{Takeshi Chiba}%
\address{
Department of Physics, College of Humanities and Sciences, \\
Nihon University, 
Tokyo 156-8550, Japan}

\date{\today}


\begin{abstract}
Recently, a model of inflation using non-minimally coupled massive vector fields 
has been proposed. For a particular choice of non-minimal coupling parameter and for a flat 
FRW model, the model is reduced to the model of chaotic inflation with massive scalar field. 
We study the effect of non-zero curvature of the universe on the onset of vector inflation. 
We find that in a curved universe the dynamics of vector inflation can be different 
from chaotic inflation, and the fraction of the initial conditions leading 
to inflationary solutions is reduced compared with the chaotic inflation case.
\end{abstract}

\maketitle

\section{Introduction}

The inflationary universe scenario has now become the standard paradigm of the early universe 
because of its success in explaining the present state of the Universe. 
Currently all successful models of inflation are based on weakly interacting scalar field(s) 
because higher spin fields generically induce a spatial  anisotropy and the effective masses  
of such fields are usually of the order of the Hubble scale and the slow-roll inflation does 
not occur \cite{ford} (see \cite{mota} for recent attempts). Therefore, it comes as a surprise 
that slow-roll inflation can be 
made possible even for massive vector fields \cite{vector} by introducing a non-minimal coupling 
of the vector fields to curvature (for the density perturbations from vector inflation, 
see \cite{kd}). The most intriguing point of the model proposed in \cite{vector} is 
that non-minimally coupled vector fields appear to behave in precisely the same way 
as a massive minimally coupled scalar field (chaotic inflation model) in a flat universe. 
Then, an immediate question is, to what extent the correspondence is perfect. 
In this paper, we address the problem from the point of view of the initial conditions for  
vector inflation as a primary (primordial) inflation, namely, how generic is vector inflation 
among all possible classical initial conditions? This problem has been  extensively 
studied for chaotic inflation \cite{belinsky,prob,piran}, and it was found that an initial 
kinetic energy or a spatial curvature of the universe generally does not prevent the onset 
of chaotic inflation and hence the inflation is general properties of the system 
(see however \cite{ghs} for the problem of the measure). 
We shall study the "vector-counterpart" of the problem studied for chaotic inflation 
in \cite{belinsky,prob}.

\section{Vector Inflation}

We briefly introduce the model of vector inflation proposed in \cite{vector}. 
The model is a massive vector field (Proca field \cite{stueck}) $U_{\mu}$ non-minimally coupled to gravity 
, and the action is given by
\beqa
S=\int \sqrt{-g}d^4x\lmk{1\over 16\pi G}R-{1\over 4}F_{\mu\nu}F^{\mu\nu}-
{1\over 2}m^2U_{\mu}U^{\mu}+{1\over 2}\xi RU_{\mu}U^{\mu}
\rmk,
\eeqa
where $F_{\mu\nu}=\partial_{\mu}U_{\nu}-\partial_{\nu}U_{\mu}$ and 
$\xi$ is a dimensionless parameter for non-minimal coupling 
and we adopt the metric signature of $(-+++)$. The correspondence between vector 
inflation and scalar inflation is given in Table \ref{tab1}. Note the difference in  
the choice of sign for $\xi$ between the vector field and scalar field.

The equations of motion are given by
\beqa
R_{\mu\nu}-{1\over 2}Rg_{\mu\nu}=&&8\pi G\lkk F_{\mu\alpha}{F_{\nu}}^{\alpha}-
{1\over 4}g_{\mu\nu}F_{\alpha\beta}F^{\alpha\beta}+  
(m^2-\xi R)U_{\mu}U_{\nu} -{1\over 2}g_{\mu\nu}(m^2-\xi R)U_{\alpha}U^{\alpha} 
\right. 
\label{eq:g}
\nonumber\\
&&
-\xi R_{\mu\nu}U_{\alpha}U^{\alpha}
+\xi (\nabla_{\mu}\nabla_{\nu}-g_{\mu\nu}\Box)U_{\alpha}U^{\alpha}
\biggr] ,\\
\nabla_{\nu}F^{\nu\mu}-m^2U^{\mu}&&+\xi RU^{\mu}=0.
\label{eq:v}
\eeqa
In particular, the equations of motions for the 
vector field in a FRW universe model are given by
\beqa
{1\over a^2}\Delta U_0 -{1\over a^2}\partial_i\dot U_i
-m^2U_0+\xi RU_0=0,\label{eq:vec1}\\
\ddot U_i+{\dot a\over a}\dot U_i-{1\over a}(a\partial_iU_0)^{.}+{1\over a^2}
\left(\partial_i(\partial_k U_k)-\Delta U_i\right)+m^2U_i-\xi RU_i=0,
\label{eq:vec2}
\eeqa
where $a$ is the scale factor, $\Delta$ is the Laplacian with respect to the spatial metric,  
the dot denotes the derivative with respect to the cosmic time and 
the summation over repeated spatial indices is assumed. 
Thus for the homogeneous vector field, Eq.(\ref{eq:vec1}) implies $U_0$=0 and, from 
Eq.(\ref{eq:vec2}), in terms of $\vp_i=U_i/a$, we obtain
\beqa
\ddot \vp_i +3H\dot \vp_i+\left(m^2+(1-6\xi)(\dot H+2H^2) -{6\xi k\over a^2}\right)\vp_i=0,
\eeqa
where we have used $R=6(\dot H+2H^2 +k/a^2)$ with $H=\dot a/a$ and $k(=0,\pm 1)$ being the spatial curvature. 
Therefore, surprisingly,  for $\xi=1/6$ \footnote{Note that the value of $\xi$ of the 
conformal coupling for a scalar field is $\xi=1/6$ in the notation given in Table \ref{tab1}.}
and for a flat universe ($k=0$), the equation of motion for 
$\vp_i$ (the norm of the vector field $U_i$) is reduced to that of minimally coupled 
massive scalar fields. 

\begin{table}
  \begin{center}
  \setlength{\tabcolsep}{3pt}
  \begin{tabular}{c|c| c|} 
   & vector & scalar \\ \hline
 Lagrangian density & $-{1\over 4}F_{\mu\nu}F^{\mu\nu}-
{1\over 2}m^2U_{\mu}U^{\mu}+{1\over 2}\xi RU_{\mu}U^{\mu}$ & 
$-{1\over 2}\nabla_{\mu}\phi\nabla^{\mu}\phi-V(\phi)-{1\over 2}\xi R\phi^2$
 \\ 
 conformal coupling & $\xi=0~~(m=0)$&  $\xi={1\over 6}~~(V=0)$ \\  
  \end{tabular}
  \end{center}
\caption{Non-minimal vector field vs. non-minimal scalar field}
\label{tab1}
\end{table}

Generally, a dynamical vector field has a preferred direction, and introducing such 
a vector field may not be consistent with the isotropy of the universe. 
In fact, the energy momentum tensor of the vector field $U_{\mu}$ (RHS of Eq.(\ref{eq:g})) 
has anisotropic components. However, 
the anisotropic part of the energy momentum tensor can be eliminated by introducing 
a triplet of mutually orthogonal vector fields  \cite{arm}. After doing that, 
we obtain the energy density $\rho$ and the pressure $p$ of the vector fields 
\beqa
\rho={3\over 2}\dot \vp_i^2+{3\over 2}m^2\vp_i^2-{3k\over 2a^2}\vp_i^2,\label{rho}\\
p={3\over 2}\dot \vp_i^2-{3\over 2}m^2\vp_i^2+{k\over 2a^2}\vp_i^2.
\eeqa
Then we finally obtain the basic equations of motion for a general FRW universe
\beqa
&&\ddot \vp_i+3H\dot \vp_i + \left(m^2-{k\over a^2}\right)\vp_i=0,\label{eq:1}\\
&&H^2+{k\over a^2}={4\pi G}\left(\dot \vp_i^2+m^2\vp_i^2-{k\over a^2}\vp_i^2\right),\label{eq:2}\\
&&\dot H+H^2=-{4\pi G}(2\dot \vp_i^2-m^2\vp_i^2),\label{eq:3}\\
&&\dot a=Ha\label{eq:4}.
\eeqa
It is noted that the effective mass squared of the vector field becomes negative and 
the field becomes tachyonic for a closed universe with $k/a^2>m^2$. 

\section{Initial Conditions for Vector Inflation}

Having introduced the basic equations, we have to specify the initial conditions for 
$a,H,\vp_i,\dot \vp_i$ in order to solve them. 
Because of the resemblance to chaotic inflation, as a primordial inflation, 
the natural choice would be the chaotic initial conditions: 
the energy density of the fields is of order of the Planck scale and one may choose arbitrary 
initial values of $\vp_i$ and $\dot \vp_i$ among all the possible choices. For the  chaotic inflation 
it has been shown \cite{belinsky,prob} that almost all of the initial conditions lead to the inflationary 
solutions in flat and open universes and in a closed universe "the probability of inflation" \footnote{There is caveat here. 
In \cite{belinsky,prob} the uniform measure in $(\phi,\dot\phi)$ space is assumed. However, 
it is not dynamically invariant measure. The natural measure contains the factor $a^3$, which leads to 
the opposite conclusion that the probability of inflation is suppressed by $\exp(-3N)$ with $N$ being 
the e-folding number \cite{ghs}.  We will not pursue this problem here. Rather we compute the 
probability of vector inflation in comparison with chaotic inflation. } 
is about 67\% for a realistic value of $m$ \cite{prob}.  
Although the universe might be inhomogeneous at the Planck epoch, as a first step toward more 
general universes, the classical evolution in a FRW universe is studied. 

However, the situation becomes ambiguous for vector inflation. 
It is not clear to what energy density we should assign the initial conditions. 
Of course one choice would be $\rho$, Eq.(\ref{rho}), derived from the energy momentum 
tensor which,  together with $p$, satisfies the energy momentum conservation; 
$\dot\rho+3H(\rho+p)=0$, and 
one may specify $\rho^{init}\sim M_{\rm pl}^4$ with $M_{\rm pl}=G^{-1/2}$ (Case (1)). 
However, it contains a curvature induced term, and such a geometric part may alternatively 
be included in the definition of the effective Planck mass, 
$M_{\rm pl~ eff}^2=M_{\rm pl}^2+4\pi \vp_i^2$. Then another possible choice would be 
$\rho_{\vp}=(3/2)(\dot \vp_i^2+m^2\vp_i^2)$ and taking $\rho_{\vp}^{init} \sim M_{\rm pl~ eff}^4$ (Case (2)). 
However, it is noted that $\rho_{\vp}$ together with $p_{\vp}=(3/2)(\dot \vp_i^2-m^2\vp_i^2)$ 
no longer satisfies the conservation law. As a more geometrical choice of the initial 
conditions, one may also consider the case of $R\sim M_{\rm pl}^2$ (Case (3)). 

One immediately find that for the first choice (Case (1)) the surface of the initial conditions 
in $(\vp_i,\dot \vp_i)$ plane can becomes {\it hyperbolic} for a closed universe with $k/a^2>m^2$
\footnote{Note that the effective mass of the vector  field is tachyonic for $k/a^2>m^2$.},  
while elliptic both in the 
latter case and in the case of chaotic inflation.  Then the possible initial values of 
the vector fields are unbounded and the initial kinetic energy of $\vp_i$ exceeds the Planck scale; 
$\dot \vp_i^2\simg M_{\rm pl}^4$, which suggests that inflation may not take place or simply 
implies that such initial conditions should be eliminated from chaotic initial conditions since 
the kinetic energy exceeds the Planck scale and the analysis cannot be treated classically. 
Restricting to the elliptic initial data for a closed universe $(k/a^2<m^2)$, then from Eq.(\ref{eq:2}) 
the allowed initial value of the expansion rate is restricted to being within the very narrow range; 
$M_{\rm pl}^2 \simg H_{init}^2\simg M_{\rm pl}^2 -m^2$. Thus from the comparison with 
chaotic inflation, we expect that the probability of vector inflation in a closed universe will be 
reduced by a factor of $ m^2/M_{\rm pl}^2$. Moreover, for an open universe, the surface of the initial 
conditions is shrunk in the $\vp_i$-direction, $\vp_i\siml M_{\rm pl}^2/\sqrt{m^2-k/a^2}$,  
and the slow-roll condition ($\vp_i\simg M_{\rm pl}$) can be violated from the beginning if 
$-k/a^2\simg M_{\rm pl}^2-m^2$. 

On the other hand, for the second choice (Case (2)), the surface of the initial conditions 
in $(\vp_i,\dot \vp_i)$ plane becomes hyperbolic irrespective of the spatial curvature of 
the universe and the kinetic energy is bounded from below: $\dot\vp_i^2\simg M_{\rm pl}^4$, 
which suggests that inflation does not occur.  

Finally, the third choice (Case (3)), using the Einstein equation, 
corresponds to $8\pi G(\rho-3p)\sim M_{\rm pl}^2$, which implies, in terms of $\vp_i$ and $\dot\vp_i$,  
$-\dot\vp_i^2+(2m^2 -k/a^2)\vp_i^2\sim M_{\rm pl}^4$. Again the surface of the initial conditions 
in $(\vp_i,\dot \vp_i)$ plane becomes hyperbolic irrespective of the spatial curvature of 
the universe and this time the potential energy has a minimum, 
$\vp_i^2\simg M_{\rm pl}^4/(2m^2-k/a^2)$, and generically dominates over 
the kinetic energy, which suggests that inflation may generically occur. Note that this applies 
also to chaotic inflation. However, for vector inflation, a closed universe with a large spatial 
curvature of $k/a^2>2m^2$ is forbidden as an initial condition. Moreover, for an open universe,  
the minimum of $\vp_i$ can be smaller than $M_{\rm pl}$ if $-k/a^2\simg M_{\rm pl}^2-2m^2$ 
and consequently some of the initial 
conditions violate the slow-roll condition from the beginning. 

In the following we consider each possibility in detail by numerically solving 
the equations of motion. 
For later convenience, we introduce the following dimensionless variables:
\beqa
\vp_i={M_{\rm pl}\over \sqrt{4\pi}}x,~~~~\dot \vp_i={mM_{\rm pl}\over \sqrt{4\pi}} y,
~~~~H=mz,~~~~t={\tau\over m},
\eeqa
where $M_{\rm pl}=G^{-1/2}$. In terms of these variables, basic equations 
Eq.(\ref{eq:1})-Eq.(\ref{eq:4}) are rewritten in the form
\beqa
&&x'=y,\label{eq:non:1}\\
&&y'=-3yz-x+{k\over a^2m^2}x,\label{eq:non:2}\\
&&z'=x^2-2y^2-z^2,\label{eq:non:3}\\
&&x^2\left(1-{k\over a^2m^2}\right)+y^2-z^2={k\over a^2m^2},\label{eq:non:4}\\
&&a'=za,\label{eq:non:5}
\eeqa
where the prime denotes the derivative with respect to $\tau$. The evolution of the universe 
is described by the trajectories in the three dimensional phase space $(x,y,z)$.

\subsection{Case (1). }

\begin{figure}
\includegraphics[width=13cm]{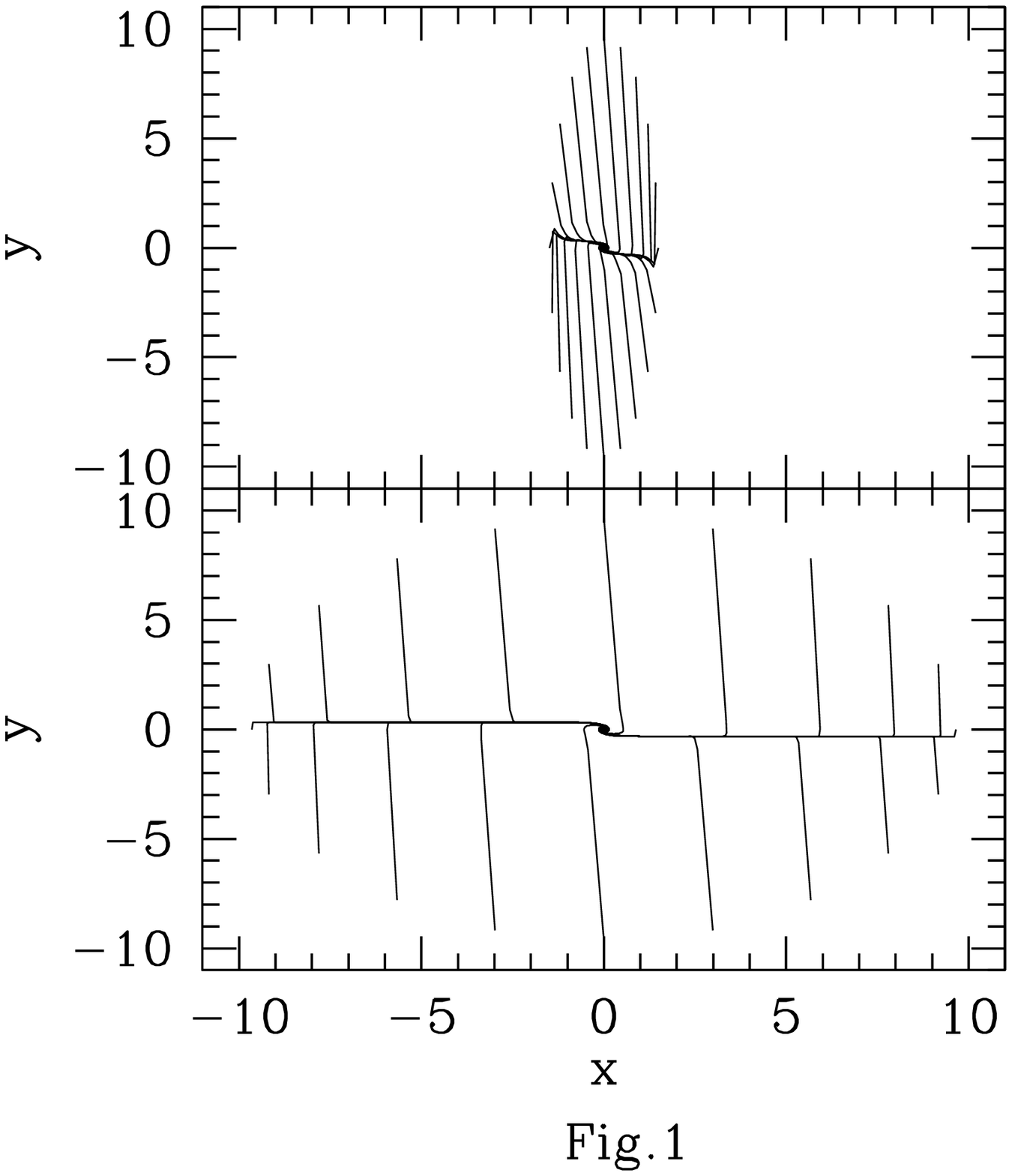}
\caption{The evolution of fields in phase space in an open universe. 
$z=1.2 M_{\rm pl}/m$ and $m=0.1M_{\rm pl}$.  The upper panel shows the result 
of vector inflation, while the lower shows the result of chaotic inflation \cite{prob}. }
\label{fig1}
\end{figure}

As a possible choice of the initial conditions for $\vp$ and $\dot\vp$, we consider
\beqa
\rho={3\over 2}(\dot \vp_i^2+m^2\vp_i^2-{k\over a^2}\vp_i^2)\sim  M_{\rm pl}^4.
\eeqa
In terms of dimensionless variable $x,y$, we set the initial conditions as
\beqa
x_{init}^2\left(1-{k\over a^2m^2}\right)+y_{init}^2=\left({M_{\rm pl}\over m}\right)^2.
\label{initial}
\eeqa
Given the initial values of $x_{init}$ and $y_{init}$, from Eq.(\ref{eq:non:4}) the initial $z_{init}$ is restricted as 
$z_{init}> M_{\rm pl}/m$ for  $k<0$ or $z_{init}< M_{\rm pl}/m$ for  $k>0$. Given $z_{init}$, 
the scale factor $a$ is determined from the constraint Eq.(\ref{eq:non:4}). 

In an open universe $(k<0$), from Eq.(\ref{initial}) the surface of the initial conditions 
in $(x,y)$ plane is an ellipse and the trajectories are inside a cone $z^2> x^2+y^2$. 
The typical trajectories projected onto the $(x,y)$ plane are shown in Fig. \ref{fig1}. 
We take $m=0.1M_{\rm pl}$ for clarity of the figures. 
The results of chaotic inflation are also shown there for 
comparison. The surface of the initial conditions is a circle then. The trajectories of the 
vector field are shrunk in the $x$-direction. Hence in $(x,y)$ plane, 
the fraction of the initial conditions leading to inflationary solutions is slightly reduced. 
Moreover, since the slow-roll  condition is violated for $\vp_i\siml M_{\rm pl}$ (or $x\siml 1$), 
{}from Eq.(\ref{initial}) vector inflation does not take place even in an open universe for 
$-k/a^2\simg M_{\rm pl}^2-m^2$. Therefore, from Eq.(\ref{eq:non:4}) and Eq.(\ref{initial}), 
the allowed range of $z_{init}$ is bounded
\beqa
z_{init}\siml \sqrt{2\left({M_{\rm pl}\over m}\right)^2-1},
\eeqa
which is in sharp contrast with chaotic inflation where the range of 
$z_{init}$ for an open universe is unbound. 

\begin{figure}
\includegraphics[width=13cm]{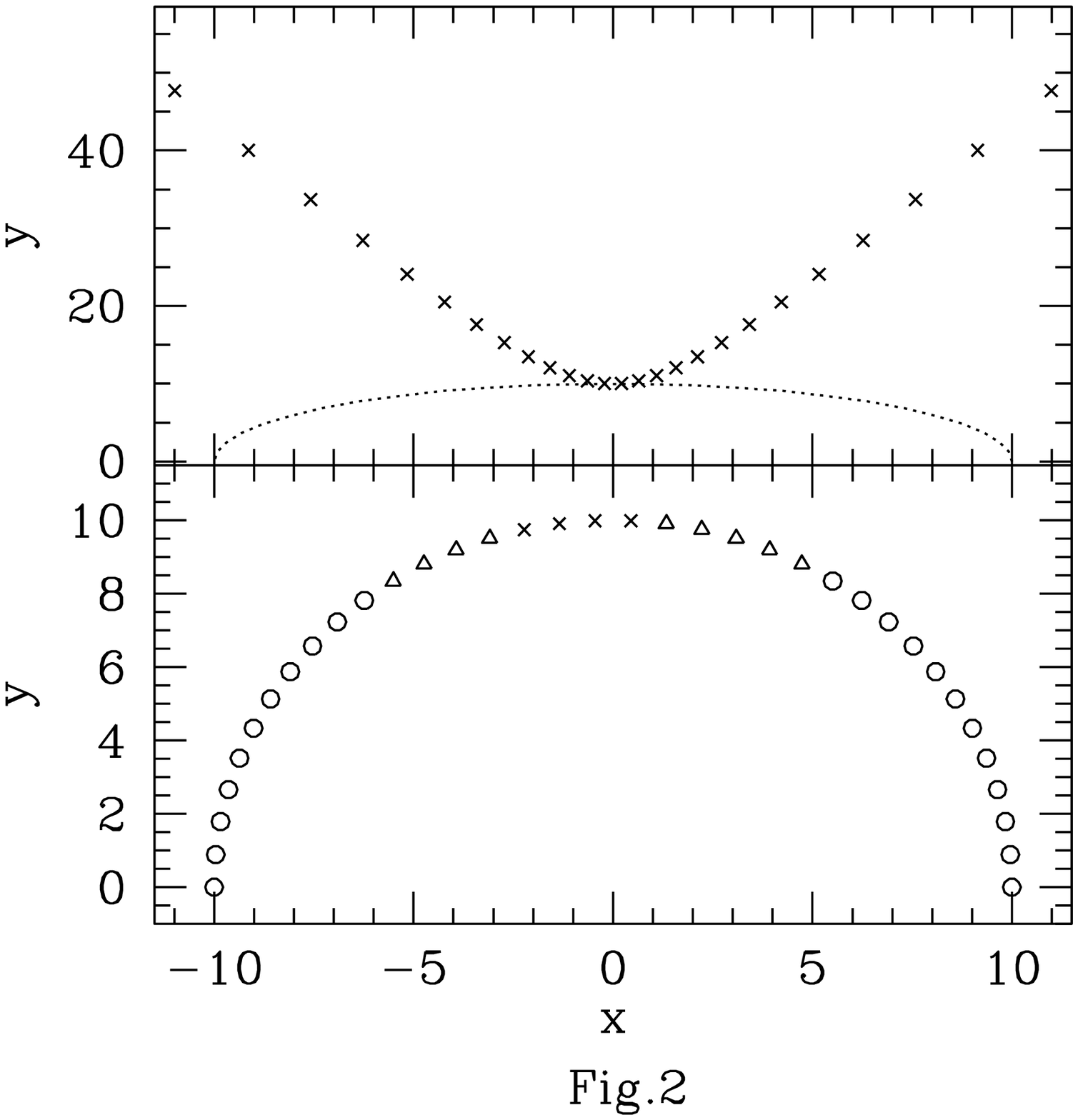}
\caption{The result of evolution from the initial conditions (Case(1)) in a closed universe with 
$z=0.9 M_{\rm pl}/m$. The circles correspond to the initial conditions which result in inflation; 
the triangles for e-folds less than 60; the crosses for recollapse. 
$m=0.1M_{\rm pl}$.  The upper (lower) panel is for vector (chaotic) inflation. The dotted curve is 
the surface of the initial conditions (quantum boundary) for chaotic inflation. }
\label{fig2}
\end{figure}

\begin{figure}
\includegraphics[width=13cm]{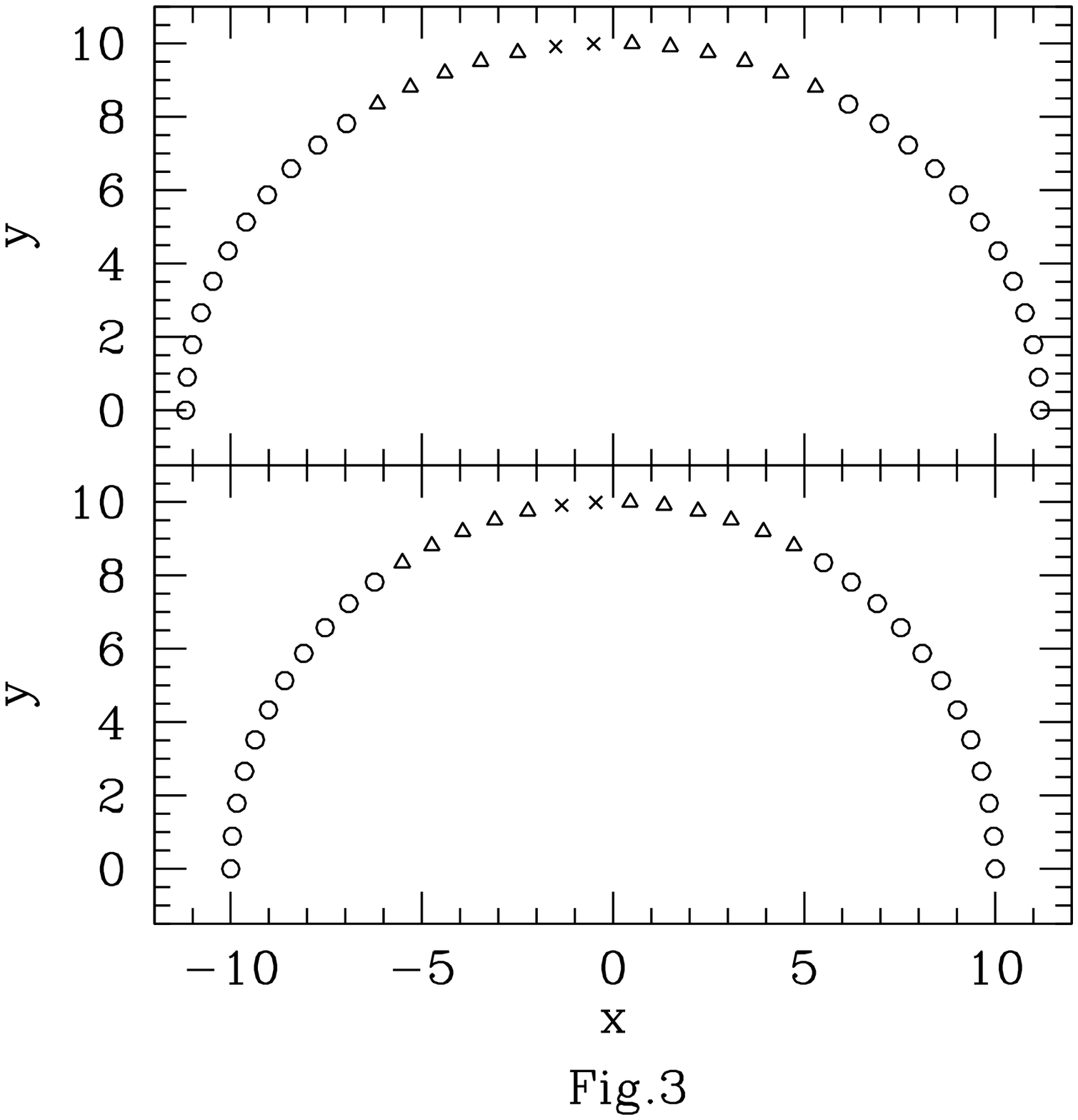}
\caption{The result of evolution from the initial conditions 
in a closed universe with $z=\left(1-0.1(m/M_{\rm pl})^2\right) M_{\rm pl}/m$. The circles correspond to 
the initial conditions which result in inflation; 
the triangles for e-folds less than 60; the crosses for recollapse. 
$m=0.1M_{\rm pl}$.  The upper (lower) panel is for vector (chaotic) inflation. }
\label{fig3}
\end{figure}

In a closed universe ($k>0$), from Eq.(\ref{initial}), 
the surface of the initial conditions in $(x,y)$ plane is an ellipse for $k/a^2<m^2$, while 
it becomes a hyperbola for $k/a^2>m^2$. Thus for $k/a^2>m^2$ the allowed values of $x$ and $y$ are unbounded, and 
moreover the initial kinetic energy already exceeds the Planck density, $y^2>(M_{\rm pl}/m)^2$, 
which suggests that inflation does not occur. 
On the other hand, for $k/a^2<m^2$, from Eq.(\ref{eq:non:4}) and Eq.(\ref{initial}), 
the allowed range of $z_{init}$ is very narrow: 
\beq
\sqrt{\left({M_{\rm pl}\over m}\right)^2-1}<z_{init}<{M_{\rm pl}\over m}.
\label{initz}
\eeq
We calculate the evolution equations Eqs.(\ref{eq:non:1}-\ref{eq:non:5}) numerically and 
determine whether inflation with enough e-folding number ($>60$) occurs or not. 
The results are given in Fig. \ref{fig2} for the case of $k/a^2>m^2$ ($z_{init}=0.9M_{\rm pl}/m$) 
and Fig. \ref{fig3} for the case of $k/a^2<m^2$ ($z_{init}=1-0.1(m/M_{\rm pl})^2)(M_{\rm pl}/m)$). 
We find that vector inflation does not occur for $k/a^2<m^2$ 
(or $z_{init}<\sqrt{(M_{\rm pl}/m)^2-1}$). Therefore the initial condition of $z$ for vector inflation 
is required to satisfy Eq.(\ref{initz}), and the probability of vector inflation is much reduced: 
$\sim (m/M_{\rm pl})^2\sim 10^{-12}$ instead of $0.67$ for $m=10^{-6}M_{\rm pl}$. 

\subsection{Case (2). }

\begin{figure}
\includegraphics[width=13cm]{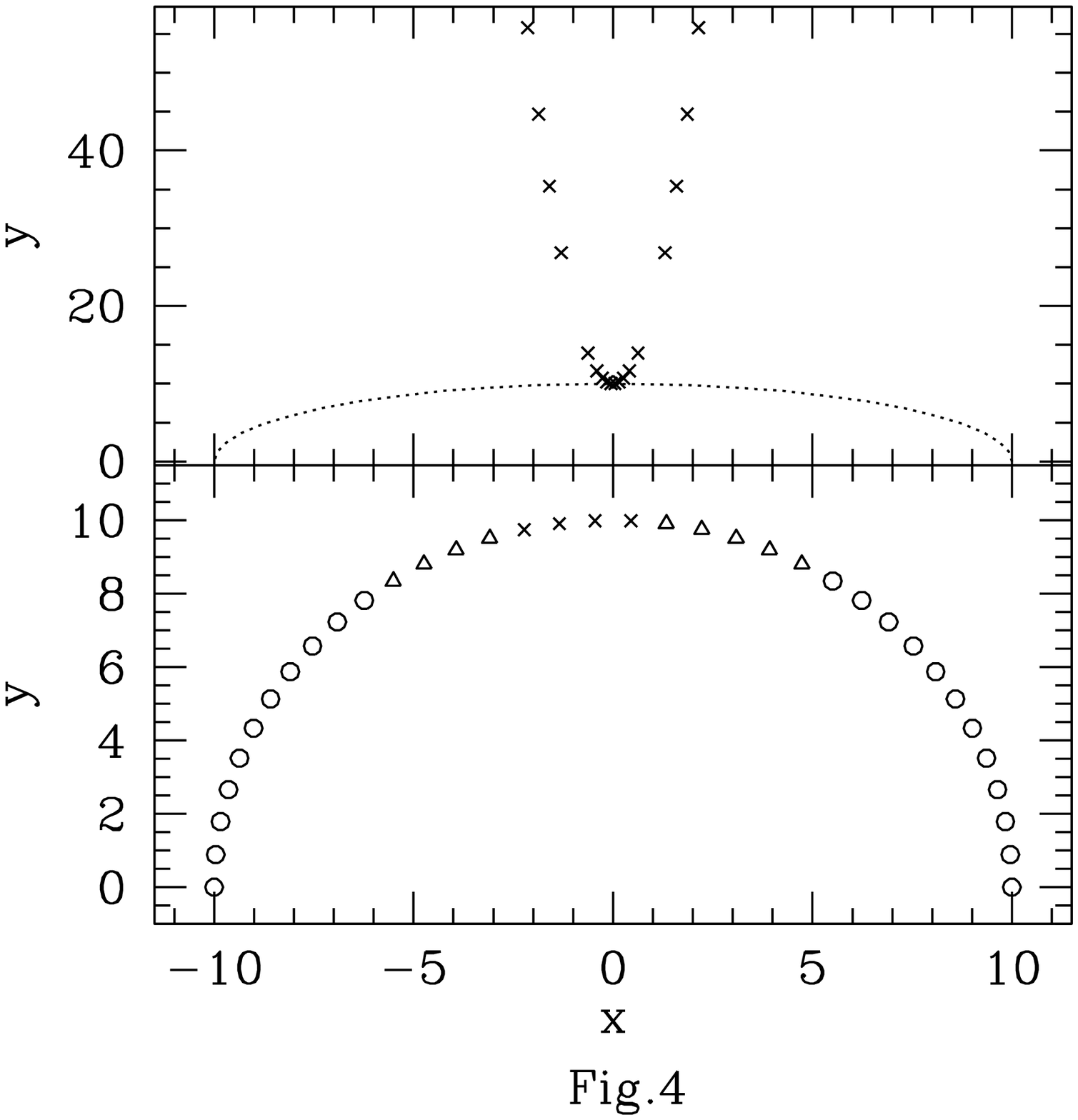}
\caption{The result of evolution from the initial conditions (Case (2)) in a closed universe with 
$z=0.9 M_{\rm pl}/m$. $m=0.1M_{\rm pl}$.  The upper (lower) panel is for vector (chaotic) inflation. 
The dotted curve is the quantum boundary for chaotic inflation. }
\label{fig4}
\end{figure}

As another choice of the initial condition, we consider
\beqa
\rho_{\vp}={3\over 2}(\dot \vp_i^2+m^2\vp_i^2)\sim  M_{\rm pl~ eff}^4=M_{\rm pl}^4
\left(1+4\pi \frac{\vp_i^2}{M_{\rm pl}^2}\right)^2.
\eeqa
Although this definition of $\rho_{\vp}$ is not derived from a covariantly conserved 
energy momentum tensor, we consider this possibility briefly just for reference. 
In terms of dimensionless variable $x,y$, we set the initial conditions as
\beqa
x_{init}^2+y_{init}^2=\left({M_{\rm pl}\over m}\right)^2(1+x_{init}^2)^2.\label{initial2}
\eeqa
Given the initial values of $x_{init}$ and $y_{init}$, from Eq.(\ref{eq:non:4}) 
the initial $z_{init}$ is restricted as 
$z_{init}> (M_{\rm pl}/m)(1+x_{init}^2)$ for  $k<0$ or $z_{init}< (M_{\rm pl}/m)(1+x_{init}^2)$ 
for  $k>0$. Given $z_{init}$, the scale factor $a$ is determined from 
the constraint Eq.(\ref{eq:non:4}). 
In this case, the surface of the initial conditions in $(x,y)$ plane becomes hyperbolic 
irrespective of $k$, and the initial kinetic energy exceeds the Planck density, 
$y^2>(M_{\rm pl}/m)^2$, In fact, from the numerical calculations, we find that vector inflation 
does not occur at all for this choice of the initial conditions. In Fig. \ref{fig4}, the results of 
the numerical integration of the evolution equations are shown for a closed universe with 
$z=0.9M_{\rm pl}/m$ for comparison with Fig. \ref{fig2}. Hence, we conclude that this choice of 
initial conditions is not appropriate for vector inflation.

\subsection{Case (3). }

As a geometrical choice of the initial condition, we consider
\beqa
R=8\pi G(\rho-3p)=24\pi G\left(-\dot\vp_i^2+2m^2\vp_i^2-{k\over a^2}\vp_i^2\right)\sim M_{\rm pl}^2.
\eeqa
In terms of dimensionless variable $x,y$, we set the initial conditions as
\beqa
x_{init}^2\left(2-{k\over a^2m^2}\right)-y_{init}^2=\left({M_{\rm pl}\over m}\right)^2.
\label{initial3}
\eeqa
Unlike the previous two cases, we first specify the initial scale factor $a$ and then specify 
$x_{init}$ and $y_{init}$ according to Eq.(\ref{initial3}) since we cannot take a $z_{init}$ which 
is uniform about $x_{init}$ and $y_{init}$. Given $a,x_{init}$ and $y_{init}$, 
$z_{init}$ is determined from the constraint Eq.(\ref{eq:non:4}). 
One immediately finds that for a closed universe the spatial curvature is restricted to the range 
$k/a^2<2m^2$.  The surface of the initial conditions in $(x,y)$ plane then becomes hyperbolic 
and the initial potential energy instead has a minimum, 
$x^2>(M_{\rm pl}/m)^2/(2-k/a^2m^2)$; this also applies to chaotic inflation.

\begin{figure}
\includegraphics[width=13cm]{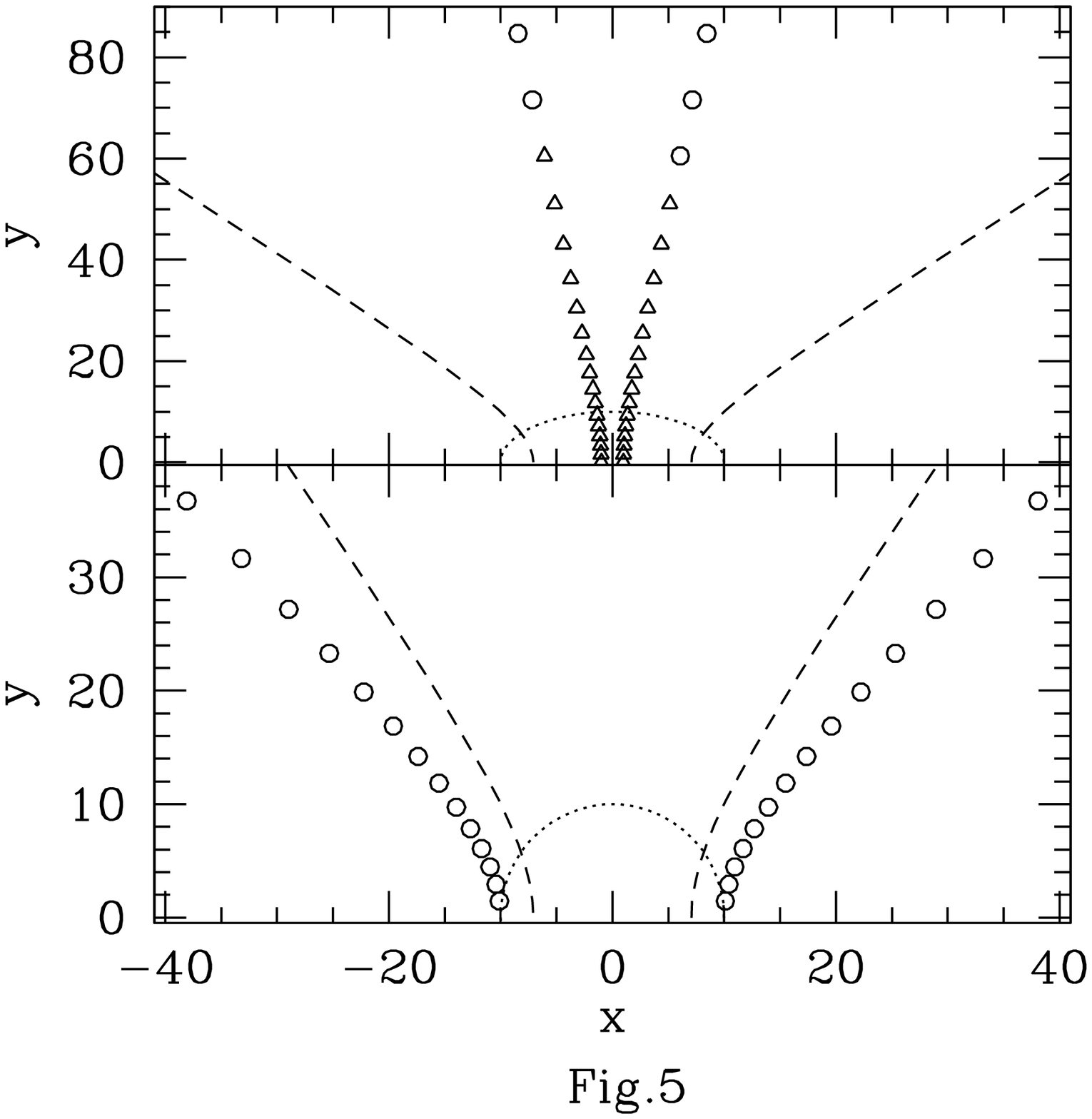}
\caption{The upper panel: The result of evolution from the initial conditions (Case (3)) 
in an open universe with $-k/a^2= M_{\rm pl}^2$.   
The lower panel: The result of evolution in a closed universe with $k/a^2= m^2$. 
$m=0.1M_{\rm pl}$.  
 Only those regions with $y>0$ are shown. The dashed curve is the surface of the same 
initial conditions for chaotic inflation. 
The dotted curve is the quantum boundary 
for chaotic inflation. }
\label{fig5}
\end{figure}

In an open universe, the minimum of $x$ can be smaller 
than unity and the slow-roll condition is violated from the beginning if 
$-k/a^2\simg M_{\rm pl}^2 -2m^2$. Then some of the initial conditions do not lead to inflationary 
solutions. In the upper half of Fig. \ref{fig5}, the results of 
the numerical integration of the evolution equations are shown for an open universe with 
$-k/a^2=M_{\rm pl}^2$, the case which violates the slow-roll conditions from the beginning. 
We find that some of the initial data do not lead to 
inflationary solutions with enough e-folds. Thus in an open universe, the initial spatial 
curvature is restricted as $-k/a^2\siml M_{\rm pl}^2 -2m^2$. 

In a closed universe, on the other hand, not all the initial data satisfying 
Eq.(\ref{initial3}) satisfy the constraint Eq.(\ref{eq:non:4}). 
{}From the positivity of $z^2$, $x_{init}$ and $y_{init}$ must satisfy
\beqa
x_{init}^2\left(1-{k\over a^2m^2}\right)+y_{init}^2>{k\over a^2m^2}.
\label{const:r}
\eeqa
The two conditions Eq.(\ref{initial3}) and Eq.(\ref{const:r}) are compatible if $k/a^2<3m^2/2$. 
Thus for a closed universe the initial spatial curvature is further constrained as $k/a^2<3m^2/2$. 
Then in a closed universe, the initial potential energy 
dominates over the kinetic energy and exceeds the Planck density from the beginning, 
$x_{init}^2>(M_{\rm pl}/m)^2/2$, which suggests that inflation always occurs. 
In fact, from the numerical calculations, we find that vector inflation does 
indeed occur as long as $k/a^2<3m^2/2$. 
In the lower half of Fig. \ref{fig5}, the results of the numerical integration of 
the evolution equations are shown for a closed universe with $k/a^2=m^2$.

\section{Summary}

The vector inflation model proposed in \cite{vector} is very similar to that of chaotic inflation. 
However, in a curved universe, the dynamics 
of vector inflation can be different from chaotic inflation.  In particular, 
in an open universe, the allowed range of the initial Hubble parameter (for Case (1)) is 
bounded. Moreover, in a closed universe, as compared with the chaotic inflation case, 
the fraction of the initial conditions (for Case (1)) which lead to inflationary solutions is much 
reduced by $(m/M_{\rm pl})^2$. 
We also find that for a naive choice of the initial conditions (Case (2)) vector inflation 
does not occur, irrespective of the spatial curvature of the universe. Further we find 
that if the initial spacetime curvature is taken to be Planckian (Case (3)), 
there are some restrictions on the initial spatial curvature, both for an open universe 
and for a closed universe which are absent for chaotic inflation 
and that the fraction of the initial conditions leading to inflationary solutions is reduced 
as compared with the chaotic inflation case. 
Since the correspondence with chaotic inflation is realized for a flat FRW universe 
and the correspondence does not hold for a general spacetime, 
it would also be interesting to study the effect of anisotropies and/or inhomogeneities on 
the onset of vector inflation along the line of \cite{moss,piran,chiba}. 
As a final remark, we note that although the dynamics of vector inflation as a 
primary (primordial) inflation is different from the dynamics of chaotic 
inflation, vector inflation as a secondary inflation (or inflation after tunneling) 
is no different from chaotic inflation since the universe is already large enough, 
and vector inflation takes place without much fine tuning of the initial conditions. 

\ack
The author would like to thank the participants of France-Japan Joint Workshop on 
the Early Universe (Nikko, Japan, 14-16 May 2008), especially Tsutomu Kobayashi, 
Hideo Kodama, Andrei Linde and Jiro Soda for useful comments and David Coule for useful 
communication. This work was supported in part by Grant-in-Aid
for Scientific Research from JSPS (No.17204018 and No.20540280)
and from MEXT (No. 20040006) and in part by Nihon University. 
The numerical calculations were performed at YITP in Kyoto University.


\end{document}